\documentclass[twocolumn,showpacs,preprintnumbers,amsmath,amssymb,aps]{revtex4}

\usepackage{amsfonts}
\usepackage[dvips]{graphicx}
\usepackage{color}

\newcommand{\eqnref}[1]{Eqn.~\eqref{#1}}
\newcommand{\figref}[1]{Fig.~\ref{#1}}
\newcommand{\secref}[1]{Sec.~\ref{#1}}
\newcommand{\tabref}[1]{Table~\ref{#1}}
\newcommand{\diffd}{\text{d}}
\newcommand{\e}[1]{\text{e}^{#1}}
\newcommand{\cmplxi}{\text{i}}
\newcommand{\bracetextsize}{\displaystyle}
\newcommand{\sgn}{\operatorname{sgn}}
\newcommand{\tr}{\operatorname{Tr}}
\renewcommand{\vec}[1]{\mathbf{#1}}
\newcommand{\obrace}[2]{\overbrace{#1}^{\bracetextsize{#2}}}

\begin{document}
\preprint{APS/123-QED}
\title{Superfluidity at the BEC-BCS crossover in two-dimensional 
Fermi gases with population and mass imbalance}
\author{G.~J. Conduit}
\email{gjc29@cam.ac.uk}
\affiliation{Theory of Condensed Matter Group, Department of Physics, 
Cavendish Laboratory, J.~J. Thomson Avenue, Cambridge, CB3 0HE, UK}
\author{P.~H. Conlon}
\affiliation{Rudolf Peierls Centre for Theoretical Physics, 1 Keble Road,
Oxford, OX1 3NP, UK}
\author{B.~D. Simons}
\affiliation{{}Theory of Condensed Matter Group, Department of Physics, 
Cavendish Laboratory, J.~J. Thomson Avenue, Cambridge, CB3 0HE, UK}
\date{\today}

\begin{abstract}
We explore the zero temperature phase behavior of a two-dimensional
two-component atomic Fermi gas with population and mass imbalance in the regime
of the BEC-BCS crossover. Working in the mean-field approximation, we show that
the normal and homogeneous balanced superfluid phases are separated by an
inhomogeneous superfluid phase of Fulde-Ferrel-Larkin-Ovchinnikov (FFLO)
type. We obtain an analytical expression for the line of continuous transitions
separating the normal and inhomogeneous FFLO phases. We further show that the
transition from the FFLO phase to the homogeneous balanced superfluid is
discontinuous leading to phase separation. If the species have different masses,
the superfluid phase is favored when the lighter species is in excess. We
explore the implications of these findings for the properties of the
two-component Fermi gas in the atomic trap geometry.  Finally, we compare and
contrast our findings with the predicted phase behavior of the electron-hole
bilayer system.
\end{abstract}


\maketitle

\section{Introduction}

By controlling interaction through a magnetically-tuned Feshbach resonance, 
ultracold atomic Fermi gases have provided a versatile arena in which to 
explore pairing phenomena and superfluidity~\cite{04rgj01,04zssrkk03,08kz01}.  
Already the crossover between the Bose-Einstein condensate (BEC) phase of 
strongly bound diatomic molecules to the BCS phase of weakly bound 
Cooper pairs has been observed experimentally~\cite{02rtbj04,03sph08,%
03ghzsdskvk06,04cbarjhg08,03grj12,04rgj01}. In recent 
years, much attention has been focused on the phase behavior of two-component
Fermi gases with population imbalance~\cite{03bcr12,05smpm07,06kjt03,06c04,%
06ps04,06pwy04,06dm05,06dorv05,06hl05,06sr05,06hs07,06lh07,06mmi09,06is09,%
07twd05,07zd10,07pwy11,07rpcbtf12,08hz01}, and 
generalized mass ratios between different species~\cite{04igotbj10,%
04szsrk10,06oshesb09,06ferzims04,06zefrim10,07gd08,07pwy11,07wsknthsgtwktj11,%
08im01}. The symmetry 
breaking effect of population and mass imbalance destabilizes the condensate
leading to an enriched phase diagram characterized by tricritical point
behavior with first order transitions separating normal and superfluid 
phases at low temperatures~\cite{07pmls02}. More detailed studies have shown 
that, on the weak coupling side of the crossover, the transition into a
homogeneous superfluid phase at low temperatures is preempted by the 
development of an inhomogeneous superfluid phase~\cite{03lw01,03bcr12,%
05mmi02,05cr08,05ss07,06sr05,06pwy04,07bhrs10,07cwzt10,07kpmbt11}. This is 
a manifestation of the FFLO phase predicted to occur in superconducting 
electron systems subject to a Zeeman field~\cite{64ff08,65lo03}. In the 
three-dimensional system, the FFLO phase is predicted to occupy only a small
region of the phase diagram making its experimental identification in the 
atomic trap geometry challenging. Indeed, even in solid state systems, the 
FFLO phase has only recently been observed~\cite{03bmcps10}.

The potential for an FFLO instability at a single wave vector in a 
three-dimensional ultracold atomic gas with only population imbalance 
was explored by \citet{06hl05} and \citet{07zd10}. They found a stable 
FFLO phase only on the BCS side of the resonance. Additionally, \citet{03wy05} 
showed the three-dimensional system is unstable to FFLO superfluid currents, 
but these were not found in the non-uniform three-dimensional trap experiments 
of \citet{06szssk07}. 

Lately, efforts have been made to explore the effects of population imbalance 
on pairing in two-component Fermi gases in two-dimensions. Although the phase
diagram of the zero temperature system has been explored in the regime of 
BEC-BCS crossover in the mean-field approximation~\cite{07twd05}, the potential
for FFLO phase formation has not yet been addressed. 
By contrast, motivated by potential applications to strongly anisotropic 
layered systems, several theoretical studies have explored the potential for 
superconducting FFLO phase formation in two-dimensional electron 
systems~\cite{95zlhr02,07pns03,07hs04}. 
A quasiclassical analysis by \citet{05cm04}, involving a 
Ginzburg-Landau expansion of the free energy in Fourier components of the 
superconducting order parameter, found that the FFLO transition in two 
dimensions was continuous at low temperatures.
In a separate study of condensation in electron-hole bilayers, \citet{07pns03} 
argued that the FFLO phase can occupy a significant part of the 
two-dimensional phase diagram. Motivated by these investigations, and the 
potential impact on the atomic gas system, in the following we will investigate
the potential for FFLO phase formation in the two-dimensional two-component
Fermi gas addressing both population imbalance and generalized mass ratios. 

In the context of ultracold atomic Fermi gases, a two-dimensional system can be
experimentally realized by confining the gas with a one-dimensional optical
lattice consisting of two counter-propagating laser beams
\cite{98ak11,01bcfmict05,01cbfmmtsi08,06cmlsssxk08}. These impose a periodic
potential, with antinodes spaced every half wavelength. The interwell barrier
energy, and therefore the tunneling rate, depends on the laser intensity, which
should be chosen to be much higher than the species chemical potential and the
superfluid gap \cite{03ms07,04wtd07}. This inhibits transfer of atoms between
layers and the Fermi gas separates into stacked quasi two-dimensional
layers. The depth of the optical potential can always be varied independently of
the external harmonic trapping potential and species chemical potentials so it
should always be possible to reduce the tunneling rate sufficiently that the
cold atom gas can be regarded as being two-dimensional gas. Within a layer, the
short-ranged interaction of the atoms can be adjusted by exploiting a Feshbach
resonance. It has been suggested \cite{08zld03} that due to the possible
formation of dressed molecules a single band theory could incorrectly predict
cloud size in the strong coupling limit. However, here we are interested in the
weak coupling limit and the behavior at the BEC-BCS crossover where we believe
that a single band theory will encompass the essential behavior allowing us to
capture the qualitative phase structure.

Finally, further motivation for the investigation of superfluidity in the 
mass imbalanced system comes from studies of exciton condensation in 
bilayer electron-hole systems. In recent years, attempts to realize a
condensed exciton phase have focused on quantum well structures where 
electrons and holes are restricted to neighboring two-dimensional
layers~\cite{95zlhr02,07pns03,07hs04}. The range of the Coulomb interaction
between the particles can be shortened by introducing a 
screening layer. As with the two-component Fermi gas, the 
electron-hole system affords the possibility of tuning between a superfluid 
of tightly-bound pairs (excitons) to a condensate phase of an electron-hole
plasma. Moreover, while one can, in principle, engineer a balanced 
electron-hole population, the effective masses of the electron and hole
quasi-particles in the semiconductor are typically quite different. In GaAs, 
the ratio of the hole to electron mass is ca. $m_{\text{h}}/m_{\text{e}}=
4.3$. Condensation phenomena in mass imbalanced systems have also been 
explored in the context of quantum chromodynamics, where the particles 
represent different species of quarks~\cite{02br09}.

The remainder of the paper is organized as follows: In 
\secref{sec:FFLOFormalism} we begin by deriving an expression
for the thermodynamic potential in the mean-field approximation for the 
two-component Fermi gas allowing for the development of an inhomogeneous 
condensate phase. In \secref{sec:FFLONonModulatedFormalsim} we use this
result to elucidate the zero-temperature phase diagram of the system for a 
uniform order parameter both at fixed chemical potential and fixed number 
density. In \secref{sec:FFLOModulatedFormalsim} we explore the tendency of the
system to condense into an inhomogeneous superfluid phase. In particular we 
combine the results of a Ginzburg-Landau expansion with the numerical 
analysis of the thermodynamic potential to infer the region over which the 
inhomogeneous phase persists. Finally, in 
\secref{sec:FFLOHarmonicallyTrappedSystem}, we examine the properties of 
the atomic Fermi gas in the harmonic trap geometry, concluding our discussion
in \secref{sec:FFLOConclusions}.

\section{Mean-field theory}\label{sec:FFLOFormalism}

Our starting point is a two-component Fermi gas with each species indexed by a
pseudo-spin $\sigma\in\{\uparrow,\downarrow\}\equiv\{+1,-1\}$. The
single-particle dispersion $\epsilon_{\vec{k},\sigma}=k^{2}/2m_{\sigma}$ depends
on the different effective masses of the two species $m_{\sigma}$, throughout we
set $\hbar=1$. Introducing the reduced mass, $1/m_{\text{R}}=(1/m_{\uparrow}+
1/m_{\downarrow})/2$ and the mass ratio $r=m_{\downarrow}/m_{\uparrow}$ we have
$m_{\uparrow}=m_{\text{R}}(1+1/r)/2$ and $m_{\downarrow}=m_{\text{R}}(1
+r)/2$. To enforce a population imbalance, each species must be associated with
its own chemical potential, $\mu_{\sigma}=\mu+\sigma h$. With these definitions,
one may see that the phase diagram of the system is symmetric under the
transformation, $(h,r)\mapsto(-h,1/r)$.

In the following, we will focus on the zero temperature phase behavior of 
the system as predicted by mean-field theory. In doing so, we will miss 
non-perturbative effects that appear at large mass ratios. In particular, 
when the ratio of masses is greater than $13.6$, it is known that, in
three-dimensions, two heavy and one light fermion can form a three-body 
weakly bound state~\cite{70e12,73e08,08bls01}. Our analysis does not include 
the possibility of such a state.

To explore the regime of BEC-BCS crossover, we will focus our attention on a
single-channel Hamiltonian describing a wide Feshbach resonance where the 
closed channel population remains small throughout~\cite{05cstl06,06yd06,%
07zd10}. The quantum partition function for the system can then be expressed
as a functional field integral over fermionic fields $\psi_{\sigma}$ and 
$\bar{\psi}_\sigma$, $\mathcal{Z}=\int\e{-S[\bar{\psi},\psi]}
\mathcal{D}\psi\mathcal{D}\bar{\psi}$, with the action
\begin{eqnarray*}
&& S\left[\psi,\bar{\psi}\right]=\int_{0}^{\beta}\diffd\tau\left[
\sum_{\vec{k},\sigma}\bar{\psi}_{\vec{k},\sigma}\left(\partial_{\tau}+
\xi_{\vec{k},\sigma}\right)\psi_{\vec{k},\sigma}\right.\nonumber\\
&&\qquad\qquad\left. -\frac{1}{2}\sum_{\vec{k},\vec{k}',\vec{q}}
\bar{\psi}_{\vec{k},\uparrow}\bar{\psi}_{\vec{q}-\vec{k},\downarrow}
V_{\vec{k}'-\vec{k}}\psi_{\vec{k}',\downarrow}\psi_{\vec{q}-\vec{k}',\uparrow}
\right]\,,
\end{eqnarray*}
where $V$ denotes the two-body interaction potential, and 
$\beta=1/k_{\text{B}}T$ is the inverse temperature. Here, for brevity, we have 
set $\xi_{\vec{k},
\sigma}=\epsilon_{\vec{k},\sigma}-\mu_{\sigma}$. Anticipating the development
of pair correlations, we introduce a Hubbard-Stratonovich decoupling of the 
interaction in the Cooper channel, with $\mathcal{Z}=\int
\e{-S[\psi,\bar{\psi},\Delta,\bar{\Delta}]}\mathcal{D}\psi\mathcal{D}
\bar{\psi}\mathcal{D}\Delta\mathcal{D}\bar{\Delta}$, where the action now
takes the form,
\begin{widetext}
\begin{eqnarray*}
&&S\left[\psi,\bar{\psi},\Delta,\bar{\Delta}\right]=\sum_{\omega,\vec{k},\vec{k}'}\bar{\Delta}_{\omega,\vec{k}}\left(V^{-1}\right)_{\vec{k}'-\vec{k}}\Delta_{\omega,\vec{k}'}\nonumber\\
&&\qquad+\sum_{\omega,\vec{k},\vec{q}}\left(\begin{array}{cc}\bar{\psi}_{\omega,\vec{q}/2-\vec{k},\uparrow}&\psi_{\omega,\vec{q}/2+\vec{k},\downarrow}\end{array}\right)\left(\begin{array}{cc}-\cmplxi\omega+\xi_{\vec{k}-\vec{q}/2,\uparrow}&\Delta_{0,\vec{q}}\\\bar{\Delta}_{0,\vec{q}}&-\cmplxi\omega-\xi_{\vec{k}+\vec{q}/2,\downarrow}\end{array}\right)\left(\begin{array}{c}\psi_{\omega,\vec{q}/2-\vec{k},\uparrow}\\\bar{\psi}_{\omega,\vec{q}/2+\vec{k},\downarrow}\end{array}\right)\,.
\end{eqnarray*}
\end{widetext}
Anticipating that the transition to the superfluid (SF) from the normal phase 
is continuous (a property already established in the weak coupling limit of 
the two-dimensional system by~\citet{05cm04}), we will suppose that the order 
parameter is characterized by a single plane-wave state corresponding to the 
stationary saddle-point solution, $\Delta_{\omega,\vec{q}}=\Delta_{\vec{Q}}
\delta_{\vec{q},\vec{Q}}\delta_{\omega,0}$~\cite{05zz10,06hl05,07zd10}. 
In this case, $\vec{Q}=\vec{0}$ 
describes the homogeneous SF state while, for $\vec{Q}\ne\vec{0}$, the 
condensate is of FFLO type. If the transition to the inhomogeneous phase
is first order, the single wave vector assumption necessitates some degree of
approximation that will underestimate the width of the FFLO region in the 
phase diagram. 

Then, approximating the functional integral over fields $\Delta$ by its 
mean-field value, and taking the interaction to be contact, $V(\vec{r})=
g\delta^2(\vec{r})$, integration over the fermionic fields gives the 
thermodynamic potential,
\begin{eqnarray}
 \label{eqn:BECBCSFFLOpreRenormalisedPotential}
&&\Omega=\frac{|\Delta_{\vec{Q}}|^{2}}{g}+\sum_{\vec{k}}(\xi_{\vec{k},+}-
\obrace{E_{\vec{k}}}{\ddagger})\nonumber\\
&&\qquad -\frac{1}{\beta}\tr\ln\left(1+\e{-\beta(E_{\vec{k}}+\sigma
\xi_{\vec{k},-})}
\right)\,,
\end{eqnarray}
where $\xi_{\vec{k},\pm}=(\xi_{\vec{k}-\vec{Q}/2,\uparrow}\pm\xi_{\vec{k}+
\vec{Q}/2,\downarrow})/2$, and $E_{\vec{k}}=(\xi_{\vec{k},+}^{2}+
|\Delta_{\vec{Q}}|^{2})^{1/2}$. From this expression, one can obtain the 
polarization or ``magnetization'',
\begin{eqnarray}
 \label{eqn:FFLOmagnetisation}
m&\equiv&n_{\uparrow}-n_{\downarrow}=-\diffd\Omega/\diffd h\nonumber\\
&=&n_{\rm F}\left(E_{\vec{k}}-\xi_{\vec{k},-}\right)-
n_{\rm F}\left(E_{\vec{k}}+\xi_{\vec{k},-}\right)\,,
\end{eqnarray}
and the total number density,
\begin{eqnarray}
 \label{eqn:FFLOnumberdensity}
&&n\equiv n_{\uparrow}+n_{\downarrow}=-\diffd\Omega/\diffd\mu\nonumber\\
&&=1+
\frac{\xi_{\vec{k},+}}{E_{\vec{k}}}\left(n_{\rm F}\left(E_{\vec{k}}-\xi_{\vec{k},-}
\right)+n_{\rm F}\left(E_{\vec{k}}+\xi_{\vec{k},-}\right)-1\right)\,,
\end{eqnarray}
where $n_{\rm F}(E)=1/(1+\text{e}^{-\beta E})$ denotes the Fermi function. 

Finally, to regularize the unphysical UV divergences associated with the 
$\delta$-function form of contact interaction (and contained within the term 
labeled by $\ddagger$ in \eqnref{eqn:BECBCSFFLOpreRenormalisedPotential}), 
we will set
\begin{equation}
 \frac{1}{g}=\int_{0}^{\infty}\frac{1}{2E+E_{\text{b}}}\diffd E\,.
\end{equation}
where $E_{\text{b}}$ denotes the energy of the two-body bound 
state~\cite{90rds01,07twd05}. $E_{\text{b}}$ will then be used as a control 
parameter to tune through the BEC-BCS crossover. As $E_{\text{b}}$ is 
increased, the system evolves continuously from the weak coupling BCS phase 
to the strong coupling BEC phase of tightly-bound pairs. 

Having obtained the thermodynamic potential in the mean-field approximation, 
we now outline our strategy for calculating the zero temperature phase 
diagram. As a platform to address the potential for inhomogeneous phase 
formation, in the following section we begin by establishing the phase 
diagram associated with a uniform order parameter, i.e. $\vec{Q}=\vec{0}$. 
In this case, the integrations associated with the thermodynamic 
potential \eqnref{eqn:BECBCSFFLOpreRenormalisedPotential} 
can be evaluated analytically and many key features of the generalized phase 
diagram understood. Then, in \secref{sec:FFLOModulatedFormalsim}, we return 
to the more general situation, exploring the capacity for inhomogeneous phase 
formation. After confirming that, in the single wave vector approximation, the 
transition to the SF phase is always continuous, we develop a Ginzburg-Landau 
expansion of the action to determine the locus of the normal-FFLO phase 
boundary analytically. Combining these results, we determine the phase diagram 
for a spatially uniform system as function of fixed chemical potential and, 
separately, as a function of fixed particle number. Finally, in 
\secref{sec:FFLOHarmonicallyTrappedSystem}, we apply these results to the
problem of resonance superfluidity in the physically realizable harmonically 
trapped system.

\section{Uniform superfluid}\label{sec:FFLONonModulatedFormalsim}

In this section we focus on the phase diagram of a system in which the 
order parameter is constrained to be uniform. At zero temperature, setting 
$\vec{Q}=\vec{0}$, the thermodynamic potential can be evaluated analytically 
for arbitrary population imbalance and mass ratio,
\begin{widetext}
 \begin{eqnarray}
  &&\frac{2\Omega}{\nu}=|\Delta_{\vec{0}}|^{2}\left[\ln\left(\frac{\sqrt{\mu^{2}+|\Delta_{\vec{0}}|^{2}}-\mu}{E_{\text{b}}}\right)-\frac{1}{2}\right]-\mu\left(\sqrt{\mu^{2}+|\Delta_{\vec{0}}|^{2}}+\mu\right)\nonumber\\
  &&-\theta\left(h'^{2}-|\Delta_{\vec{0}}|^{2}\right)\left[\theta\left(\mu_{\text{f}}-Rh'-\mu_{\text{c}}\right)\left(2|h'|\mu_{\text{c}}-|\Delta_{\vec{0}}|^{2}\ln\left|\frac{\mu_{\text{c}}+|h'|}{\mu_{\text{c}}-|h'|}\right|\right)\right.\nonumber\\
  &&+\theta\left(\mu_{\text{c}}+\mu_{\text{f}}-Rh'\right)\theta\left(\mu_{\text{c}}-\mu_{\text{f}}+Rh'\right)\nonumber\\
  &&\times\left(|h'|(\mu_{\text{c}}+2\mu_{\text{f}})-\mu\sqrt{\mu^{2}+|\Delta_{\vec{0}}|^{2}}-|\Delta_{\vec{0}}|^{2}\ln\left|\frac{\mu_{\text{c}}+|h'|}{\sqrt{\mu^{2}+|\Delta_{\vec{0}}|^{2}}-\mu}\right|\right.\nonumber\\
  &&+\left.\left.\sgn(h')\left(R(\mu^{2}-h'^{2})+\frac{|\Delta_{\vec{0}}|^{2}}{2}\ln(r)\right)\right)\right]\,.
 \end{eqnarray}
\end{widetext}
Here $\theta$ denotes the Heaviside step function, $\nu=m_{\text{R}}/2\pi$ the 
two-dimensional density of states of the reduced mass system and, for clarity, 
we have defined the set of parameters,
\begin{eqnarray}
 &&R\equiv\frac{r-1}{r+1},\qquad h'\equiv\frac{h-\mu R}{\sqrt{1-R^{2}}}
 \nonumber\\
 &&\mu_{\text{c}}\equiv\sqrt{h'^{2}-|\Delta_{\vec{0}}|^{2}},\qquad
 \mu_{\text{f}}\equiv\mu\sqrt{1-R^{2}}\,.
\end{eqnarray}
By minimizing the thermodynamic potential with respect to $\Delta_{\vec{0}}$, 
one obtains the loci of phase boundaries shown in 
\tabref{tab:FFLOAnalyticalPhaseBoundary}. When the mass ratio is unity 
($R=0$), these results coincide with those obtained in Ref.~\cite{07twd05}.
In particular, one may note that, in the SF phase, the order parameter 
takes the form 
\begin{eqnarray*}\label{eqn:OrderParameter}
 |\Delta_{\vec{0}}|=\sqrt{E_{\text{b}}(2\mu+E_{\text{b}})}\,,
\end{eqnarray*}
implying a thermodynamic potential, $\Omega=-\nu(\mu+E_{\text{b}}/2)^{2}$, 
independent of the chemical potential difference, $h$. As a result, one may 
infer that the magnetization, $m=-\diffd\Omega/\diffd h$, is zero. For the 
uniform condensate, the SF phase always involves a balanced population of 
fermions.
Drawing on these results, let us now comment on the implications for the phase
diagram of the system for, respectively, fixed chemical potential and fixed 
particle number. 

\begin{table*}
  \caption{Summary detailing the loci of phase boundaries for 
    $\mu/E_{\text{b}}$ as a function of $h/E_{\text{b}}$ and $R=(r-1)/(r+1)$. 
    Results labeled 
    ($\bigstar$) are found in, and are relevant for, 
    \secref{sec:FFLOModulatedFormalsim}.}
  \begin{tabular}{cc}
   \hline
   PP-FFLO$\bigstar$&$\displaystyle\frac{1+(h/E_{\text{b}})R\pm\sqrt{1+2(h/E_{\text{b}})R-R^{2}-2(h/E_{\text{b}})R^{3}}}{R^{2}}$\\
   FFLO-SF$\bigstar$ and PP-SF&$\displaystyle\frac{1+2(h/E_{\text{b}})R-R^{2}-\sqrt{(1-R^{2})(1+4(h/E_{\text{b}})R)}}{2R^{2}}$\\
   FP-SF&$\displaystyle\frac{\sqrt{2}(h/E_{\text{b}})+\sqrt{1-R}}{\sqrt{2}-2\sqrt{1-R}}$\\
   ZP-FP&$\displaystyle\pm(h/E_{\text{b}})$\\
   FP-FFLO$\bigstar$ and FP-PP&$\displaystyle\mp(h/E_{\text{b}})$\\
   ZP-SF&$\displaystyle-1/2$\\
   \hline
  \end{tabular}
  \label{tab:FFLOAnalyticalPhaseBoundary}
 \end{table*}

\subsection{Fixed chemical potentials}\label{sec:ConstChemPotZeroQ}

\begin{figure}
 \centerline{\resizebox{\linewidth}{!}{\includegraphics{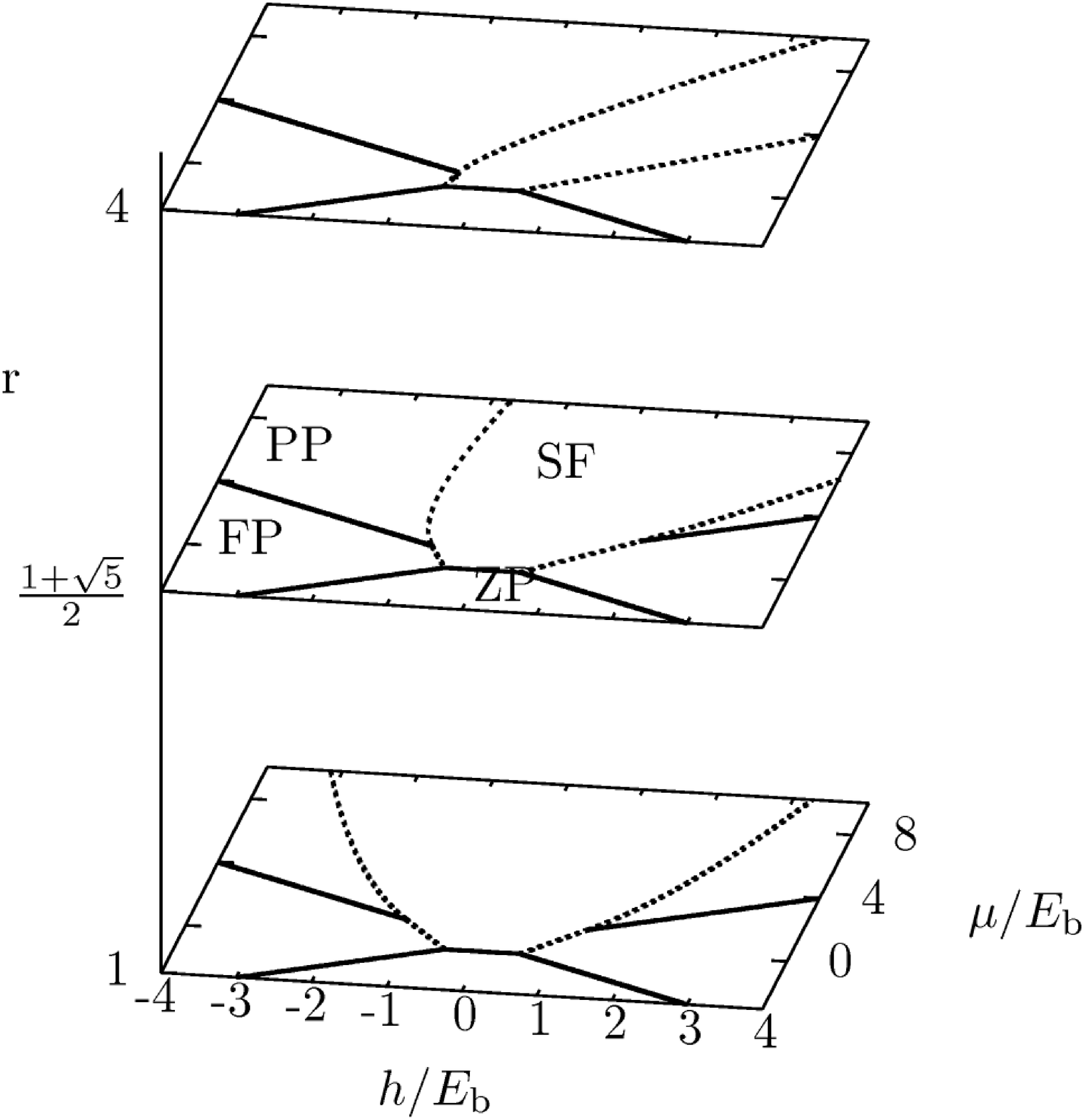}}}
 \caption{The phase diagram shown as a function of $(h/E_{\text{b}},
 \mu/E_{\text{b}})$ for three different values of the mass ratio, $r$. The 
 diagrams were constructed assuming a uniform order parameter, neglecting the 
 potential for inhomogeneous phase formation. The solid lines represent 
 continuous phase boundaries, while the dashed lines denote first order 
 transitions into the balanced SF phase.}
 \label{fig:OmegaPhaseZeroQ}
\end{figure}

When the chemical potentials, $\mu$ and $h$, are held constant, minimization 
of the thermodynamic potential leads to the phase diagram depicted in 
\figref{fig:OmegaPhaseZeroQ}. The equal mass case is consistent with the result of 
\citet{07twd05}. For $\mu$ smaller than either the molecular 
binding energy per particle, $-E_{\text{b}}/2$, or the chemical potential 
shift associated with the majority species, $-h$, (corresponding to an empty 
Fermi surface), the equilibrium phase hosts no particles (the 
``zero-particle'' state, ZP). On increasing the chemical potential, $\mu$, a 
second order phase transition into either a balanced SF, or a fully-polarized 
(FP) normal phase occurs. The transfers from the zero particle phase to a 
FP normal phase, and from a FP phase to a partially-polarized (PP) normal 
phase are both continuous. The phases have boundaries where the Fermi surface 
shrinks to zero at $\mu=-h$ and $\mu=+h$ respectively (for $h>0$). At fixed 
$E_{\text{b}}$, an increase in chemical potential, $\mu$, leads to an 
increase in the order parameter of the balanced superfluid system, $|\Delta_{0}|
\propto \sqrt{\mu}$, and an attendant increase in the critical $h$ required 
to destroy the condensate. The phase transition from the normal state, both 
FP and PP, into the SF is first order. 

As the ratio of masses is increased, as shown in \figref{fig:OmegaPhaseZeroQ} 
on the side $r>1$, the phase diagram becomes skewed. This can be understood 
by tracking the locus of the line where the Fermi surfaces of the two species 
are perfectly matched, approximately along the center of the SF phase. 
The central superfluid locus is $\mu/h=1/R$, which is consistent with the skew. 
Superfluidity is therefore more favorable if the ``light'' species has a 
greater chemical potential than the ``heavy'' species.

\subsection{Fixed number densities}\label{sec:ConstNumDensZeroQ}

\begin{figure}
 \includegraphics{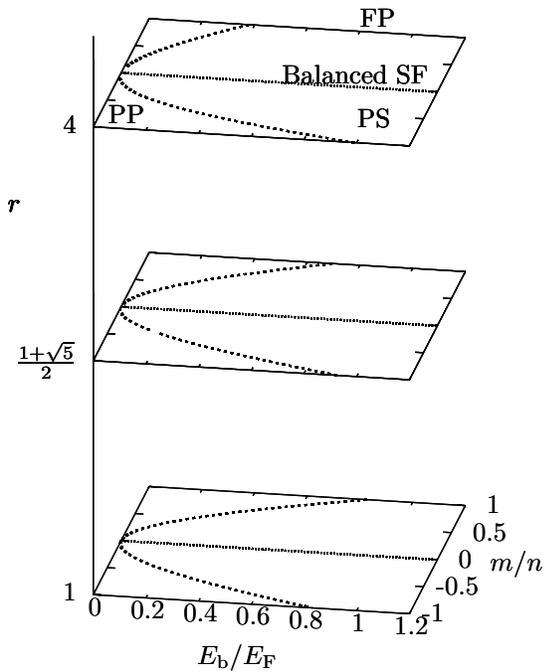}
 \caption{The phase diagram as a function of $m/n$ and interaction 
  strength, $E_{\text{b}}/E_{\text{F}}$, of the two-dimensional system with 
  fixed majority and minority particle densities for three different mass 
  ratios, $r$. The SF phase (dotted line) is the line of zero population 
  imbalance. Between the balanced SF and PP/FP phase (dashed line) lies a 
  region of phase separation (PS).}
 \label{fig:FFLOUniformFreeZeroQ}
\end{figure}

In the canonical ensemble, where the number densities $n$ and $m$ are held 
constant, the chemical potentials, $\mu$ and $h$, must be inferred 
self-consistently. In this case, a first order transition in the 
$(\mu/E_{\text{b}},h/E_{\text{b}})$ phase diagram 
(\figref{fig:OmegaPhaseZeroQ}) implies phase separation (PS)~\cite{03bcr12} 
in the $(n,m)$ phase diagram. At each point along the PP-SF phase boundary 
in $(\mu/E_{\text{b}},h/E_{\text{b}})$ one can evaluate the corresponding
polarization and total number density. From this result, one can infer
the boundaries between the normal and phase separated regions as functions 
of $E_{\text{b}}/E_{\text{F}}$ and polarization, $m/n$. Here we have defined
a ``Fermi energy'' scale $E_{\text{F}}=n/\nu$, where $\nu=m_{\text{R}}/2\pi$ 
denotes the constant
two-dimensional density of states of the reduced mass system. 
The resulting phase diagram is shown in \figref{fig:FFLOUniformFreeZeroQ}.

As expected, in the BEC limit of large $E_{\text{b}}/E_{\text{F}}$, one
finds phase separation, with the development of a condensate of tightly-bound 
molecular pairs coexisting with a FP phase containing excess fermions. The 
phase diagram shows that this behavior persists into the weak coupling BCS 
limit, with the system phase separating into a balanced SF phase (i.e. with 
$m/n=0$), and the excess particles forming a non-interacting PP Fermi gas. In 
the BCS limit of weak pairing, a small population imbalance is sufficient to 
destroy pairing and enter the PP normal phase region. 

When the species have unequal masses, the phase diagram is skewed, 
similar to the fixed chemical potential case in 
\secref{sec:ConstChemPotZeroQ}. If there is a mass 
imbalance then the Fermi energy of each spin species scales as 
$E_{\text{F},\sigma}=n_{\sigma}/\nu_{\sigma}\equiv \pi(n+\sigma m)/m_{\sigma}
\propto1/m_{\sigma}$ implying that it is energetically 
more beneficial for the ``heavy'' rather than ``light'' particles to be in 
the normal state. Therefore, at a given mass imbalance, the phase diagram 
loses its symmetry in $m/n$ and superfluidity is favored if the ``lighter'' 
species is in excess whereas the normal state is favored if the ``heavy'' 
species is in excess.

\section{Inhomogeneous superfluid}\label{sec:FFLOModulatedFormalsim}

With the properties of the uniform SF phase in place, we now turn to the 
question of inhomogeneous phase formation. To characterize the nature of the 
PP-FFLO transition, we adopt two methodologies: firstly, in 
\secref{sec:FFLOGLFormalism}, we will develop a Ginzburg-Landau expansion of 
the action to explore the locus of putative continuous transitions from the 
normal PP phase into the inhomogeneous FFLO phase. Secondly, in 
\secref{sec:FFLOTransitionOrder}, we will assess the validity of the
Ginzburg-Landau expansion by investigating the global minimum of the 
thermodynamic potential for a mean-field order parameter field involving 
a single wave vector. Using these results, we will infer the phase diagram 
of a system with fixed 
chemical potentials in \secref{sec:FFLOFreeSystem}, and fixed particle 
densities in \secref{sec:FFLOUniformTrappedSystem}.

\subsection{Ginzburg-Landau theory}\label{sec:FFLOGLFormalism}

With the Ansatz that the transition from the normal to condensed phase is 
continuous, close to the transition we may expand the action in fluctuations, 
$|\Delta_{\vec{q}}|$. In doing so, one obtains
\begin{equation}
 S_{\text{eff}}=\sum_{\vec{q}}\alpha_{\vec{q}}|\Delta_{\vec{q}}|^{2}+
\mathcal{O}\left(|\Delta|^4\right)\,,
\end{equation}
where 
\begin{eqnarray*}
\alpha_{\vec{q}}=
\sum_{\vec{k}}\left(\frac{1}{2\epsilon_{\vec{k}}+E_{\text{b}}}-\frac{1-n(\xi_{\vec{k}-\vec{q}/2,\uparrow})-n(\xi_{\vec{k}+\vec{q}/2,\downarrow})}{\xi_{\vec{k}-\vec{q}/2,\uparrow}+\xi_{\vec{k}+\vec{q}/2,\downarrow}}\right)
\end{eqnarray*}
denotes the static pair susceptibility. The locus of continuous transitions 
may be determined from the value of $\vec{q}$ at which $\alpha_{\vec{q}}$ is 
both minimized and passes through zero. Within the condensed phase, higher 
order terms in $\Delta_{\vec{q}}$ determine the crystalline structure of the 
FFLO state \cite{04cn02}. 

The corresponding phase boundary then translates to the largest allowable 
chemical potential shift, $h$, which occurs when the Fermi surfaces just touch 
but do not cross~\cite{05cm04}. From this condition, one finds a phase 
boundary along the line,
\begin{equation}
 \label{eqn:GLTheoryRecastBoundary}
 \frac{h}{E_{\text{b}}}=\left(\frac{\mu}{E_{\text{b}}}-1\right)R\pm
 \sqrt{\left(\frac{2\mu}{E_{\text{b}}}-1\right)\left(1-R^{2}\right)}\,.
\end{equation}
Minimizing $\alpha_{\vec{q}}$ with respect to $|\vec{q}|$, one obtains the further 
condition $\epsilon_{\vec{q}}\equiv q^{2}/(2m_{\text{R}})=2E_{\text{b}}/
(1-R^{2})$. Measured in units of the Fermi momentum of the reduced mass
system, this translates to a wave vector,
\begin{equation}
\frac{|\vec{q}|}{k_{\text{F}}}=\sqrt{\frac{E_{\text{b}}}{E_{\text{F}}}\frac{
(m_{\uparrow}+m_{\downarrow})}{2m_{\text{R}}}}\,,
\end{equation}
where $k_{\text{F}}^{2}=2m_{\text{R}}E_{\text{F}}$ and, inverting 
\eqnref{eqn:GLTheoryRecastBoundary}, $E_{\text{b}}=\mu-h R\pm\sqrt{(\mu^{2}-
h^{2})(1-R^{2})}$. In the weak coupling limit, $E_{\text{b}}\ll E_{\text{F}}$, 
so that, at equal masses, $E_{\text{b}}=h^{2}/2\mu$ giving $|\vec{q}|=
2h/v_{\text{F}}$, where $v_{\text{F}}$ is the Fermi velocity, agreeing with 
the findings of \citet{94br02,94s11}, and \citet{05cm04}. In the same limit,
the pair susceptibility takes the form, $\Re(\ln(1+\sqrt{1-
(|\vec{q}|v_{\text{F}}/2h)^{2}}))$, collapsing to that found in previous works.

\subsection{FFLO instability phase boundaries}\label{sec:FFLOTransitionOrder}

\begin{figure}
 \includegraphics{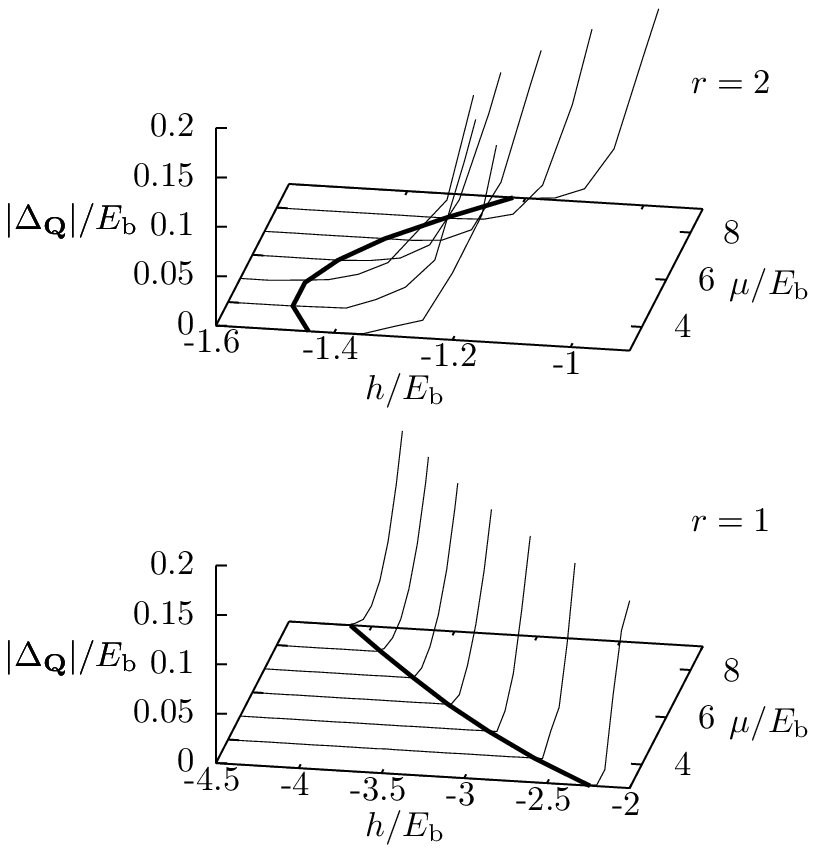}
 \caption{Variation of the order parameter $|\Delta_{\vec{Q}}|/E_{\text{b}}$ with 
chemical potential difference, $h/E_{\text{b}}$, and chemical potential, 
$\mu/E_{\text{b}}$. The upper panel is at a mass ratio $r=2$, and the 
lower panel at equal masses, $r=1$. The thin black lines trace out the 
$|\Delta_{\vec{Q}}|/E_{\text{b}}$ variation, found by minimizing the thermodynamic 
potential \eqnref{eqn:BECBCSFFLOpreRenormalisedPotential}, for several 
different chemical potentials. The thick black line is the locus of the 
second order transition predicted by Ginzburg-Landau theory.}
 \label{fig:FFLODeltaVariation}
\end{figure}

To assess whether the transition from the PP phase to the FFLO phase is really
continuous, one can instead minimize the thermodynamic potential
\eqnref{eqn:BECBCSFFLOpreRenormalisedPotential} with respect to the wave vector
$\vec{Q}$ and the mean-field value of the order parameter $\Delta_{\vec{Q}}$.
For several values of chemical potential, $\mu$, and two different mass ratios
$r=1$ and $r=2$, numerical minimization of the thermodynamic potential confirms
that the order parameter changes continuously (see
\figref{fig:FFLODeltaVariation}), falling to zero along a line of
instability. The locus of the transition also agrees with that obtained from the
Ginzburg-Landau expansion in \secref{sec:FFLOGLFormalism}. This result is in
accord with that found in Ref.~\cite{05cm04} in the weak coupling limit of the
equal mass system, and shows that the transition remains continuous across the
entire range of the FFLO phase.

We are now in a position to evaluate all phase boundaries associated with the 
FFLO instability. The agreement described above between Ginzburg-Landau theory 
and direct minimization allows us to use the analytic Ginzburg-Landau boundary
between the PP and FFLO phases. The minimum in the thermodynamic potential 
that gives rise to the FFLO phase is shallow relative to that of the SF 
phase. We are therefore able to approximate the actual FFLO-SF phase boundary 
by the $\vec{Q}=\vec{0}$ result for the PP-SF boundary described in 
\secref{sec:FFLONonModulatedFormalsim}. A summary of the phase boundaries 
is shown in \tabref{tab:FFLOAnalyticalPhaseBoundary}, the additional 
boundaries due to the presence of the FFLO phase are labeled ($\bigstar$). 
As the extent of the SF region is only reduced by the presence of the FFLO 
phase, the SF is balanced, as was shown for the $\vec{Q}=\vec{0}$ study in 
\secref{sec:FFLONonModulatedFormalsim}.

\subsection{Fixed chemical potentials}\label{sec:FFLOFreeSystem}

\begin{figure*}
 \centerline{\resizebox{\linewidth}{!}{\includegraphics{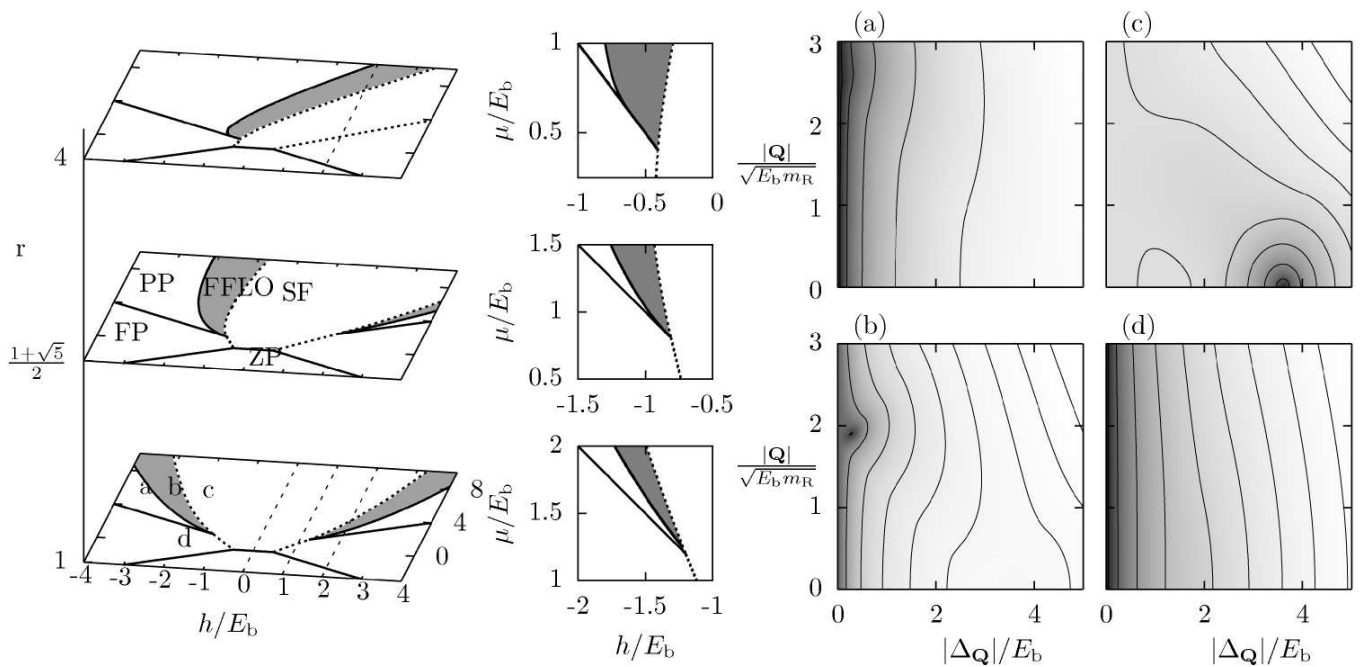}}}
 \caption{The left-hand column shows the phase diagram in 
$(\mu/E_{\text{b}},h/E_{\text{b}})$ at three 
different mass ratios $r$. The solid lines represent second order phase 
boundaries, the dashed line denotes first order phase transitions into the 
balanced SF phase. The FFLO phase is highlighted in grey. The trajectories 
followed in the sample traps in \secref{sec:FFLOHarmonicallyTrappedSystem} 
are shown by thin dashed lines. The central column of diagrams focus more
closely on the topology of the phase diagram close to the tricritical point 
region for $h/E_{\text{b}}<0$. Thermodynamic potential 
surfaces are shown in the right-hand column; the darker the more negative 
and so more favorable; label (a) is a PP normal state, (b) a FFLO state, 
(c) a SF state, and (d) a FP state. Different plots have different shading 
calibrations, (unequally spaced) contour lines are also shown.}
 \label{fig:OmegaPhase}
\end{figure*}

Let us now apply these results to the problem of a uniform system with 
constant chemical potentials. The corresponding phase diagram is shown in 
\figref{fig:OmegaPhase}. While the general topology of the phase diagram
mirrors that discussed in \secref{sec:ConstChemPotZeroQ}, the transition to
the balanced SF phase is preempted by the formation of an inhomogeneous
FFLO phase. The FFLO instability occurs mainly on the PP side of the PP-SF 
phase boundary of the uniform condensate shown in \figref{fig:OmegaPhaseZeroQ} 
with just a small intrusion on the balanced SF side. The FFLO instability 
does not occur within the FP state as there are no minority state particles 
with which to pair. The FFLO-PP boundary is second order, while the FFLO-SF 
boundary is first order.

The FFLO-PP phase boundary terminates at the SF phase for small mass ratios 
and at the FP phase for large mass ratios on the side of the majority 
``heavy'' species. The movement of the boundary terminus with increasing 
mass ratio $r$ is in the opposite direction on the majority ``light'' species 
side -- it moves further up the SF phase boundary. The special mass ratio 
where it terminates at the SF-FP phase boundary on the majority ``heavy'' 
species side is at $r_{\text{c}}=(1+\sqrt{5})/2$.

The thermodynamic potential variation is also shown in \figref{fig:OmegaPhase} 
at four different points (a, b, c, d) for $r=1$. Since the wave vector
dependence of the thermodynamic potential enters through the order parameter, 
in both the PP (a) and FP (d) normal phases the minimum is 
$|\vec{Q}|$-independent. At the highlighted FFLO phase point (b), the global 
minimum lies at $|\Delta_{\vec{Q}}|\approx0.2E_{\text{b}}$ with $|\vec{Q}|\approx
2\sqrt{m_\text{R}E_{\text{b}}}$, while a local minimum also develops at 
$|\Delta_{\vec{Q}}|\approx3.8E_{\text{b}}$ with $|\vec{Q}|=0$ corresponding to the 
putative uniform SF phase. At the highlighted SF point (c), the global 
minimum lies at $|\Delta_{\vec{Q}}|\approx3.8E_{\text{b}}$ and $|\vec{Q}|=0$.

\subsection{Fixed number densities}\label{sec:FFLOUniformTrappedSystem}

\begin{figure*}
 \includegraphics{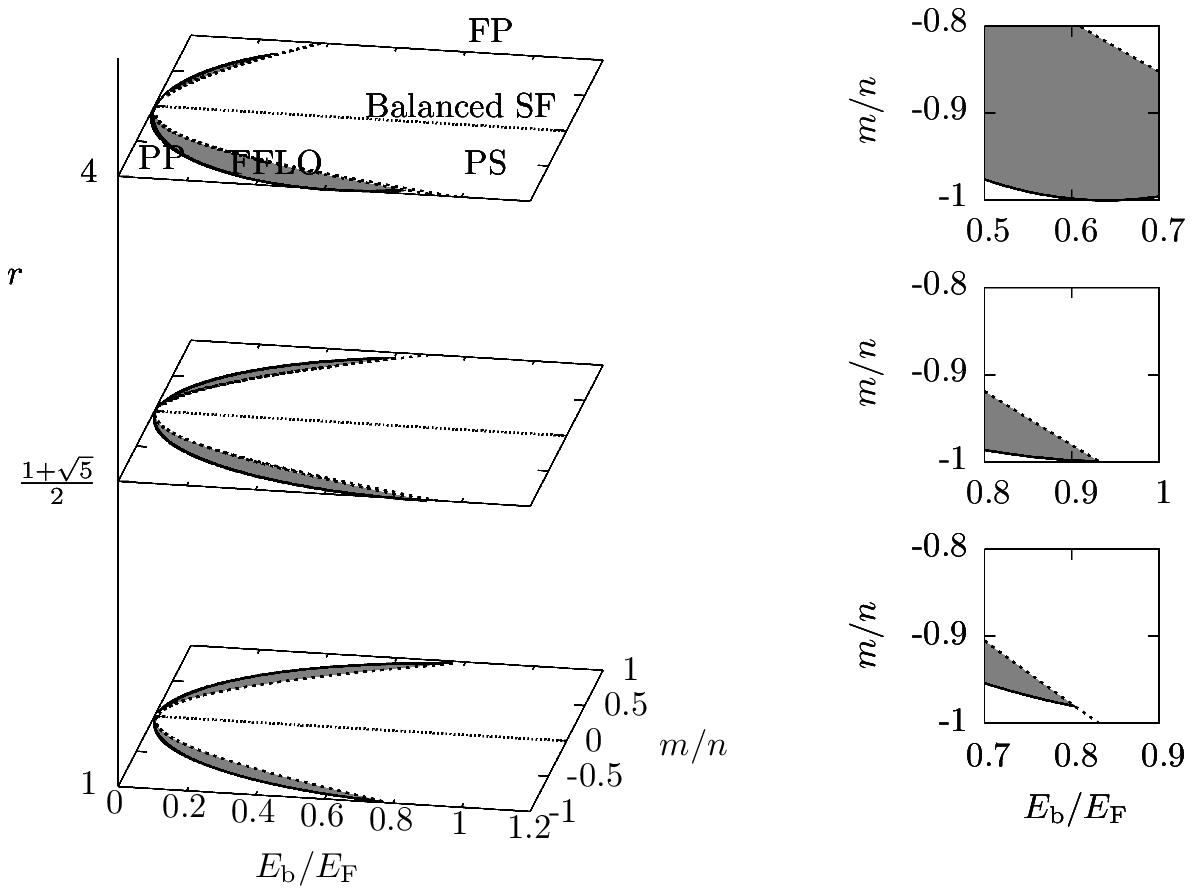}
 \caption{The phase diagram in $(m/n,E_{\text{b}}/E_{\text{F}})$ of the 
two-dimensional system with fixed majority and minority particle densities for 
three different mass ratios $r$. The SF phase (dotted line) is represented by
the line of zero population imbalance $m/n=0$. The PP phase is separated from the 
FFLO phase (highlighted in grey) by a second order phase boundary (solid 
line). In the phase separated (PS) region bordered by the dashed line, the 
system separates into a balanced superfluid (SF) and either a normal phase 
or, depending on the composition, an inhomogeneous FFLO phase. The right-hand 
column of graphs focus on the terminus of the PP-FFLO phase boundary on the 
side of negative polarization for the three featured mass ratios.}
 \label{fig:FFLOUniformFree}
\end{figure*}

Let us now address the implications of the phase diagram for a spatially 
uniform system held at fixed number densities. Obtaining the corresponding 
density, $n$, and magnetization, $m$, gives the phase diagram shown 
in \figref{fig:FFLOUniformFree}. Once again, the topology of the phase 
diagram mirrors that discussed for the homogeneous condensate in  
\secref{sec:ConstNumDensZeroQ}. However, between the phase separated SF
phase and normal phase, the system exhibits an inhomogeneous FFLO phase 
over a wide region of the phase diagram. 

In the weak coupling BCS limit, even a small population imbalance is 
sufficient to enter the FFLO phase region. We note that in a population 
balanced system the Fermi momenta 
of the populations are equal so no shift of the Fermi surfaces is required 
to form Cooper pairs and a modulated phase is not seen.

The effects of the moving PP-FFLO 
phase boundary terminus, described in \secref{sec:FFLOFreeSystem}, are also 
apparent. For equal masses, the FFLO phase never meets the FP normal 
state. For mass ratios in excess of $r_{\text{c}}=(1+\sqrt{5})/2$, the FFLO 
phase meets the FP state on the majority ``heavy'' species side, but is 
further from the FP state on the ``light'' species side. For high mass ratios, 
this is evidenced by the much broader FFLO region on the heavy species side. 

To conclude this section, it is interesting to compare 
the phase diagram of the ultracold atom system with contact 
interaction and the problem of electron-hole bilayers with long-ranged 
Coulomb interaction. In particular, we focus our discussion on the study 
in Ref.~\cite{07pns03} of GaAs bilayers where the mass ratio $r=4.3$. In
this case, it is more natural to characterize the strength of interaction 
by $r_{\text{s}}=r_0/a_{0}$, where $r_0=1/\sqrt{\pi n}$ denotes the 
interparticle spacing, and $a_0$ is the effective Bohr radius of the 
two-body bound state. The latter is related to the dimensionless
ratio $E_{\text{b}}/E_{\text{F}}$ through the relation, $E_{\text{b}}/
E_{\text{F}}=0.381r_{\text{s}}^{2}$. As a result, we find that the system 
enters the BCS phase with the appearance of FFLO phase behavior for 
$r_{\text{s}}$ values of ca. 1.5 (4) compared with that found for the 
unscreened electron-hole bilayer of $r_{\text{s}}\sim1.5$ (16) for the 
``light'' (``heavy'') species. More qualitatively, in both cases, the systems 
show a preference towards the superfluid phase when the ``light'' species is 
in excess, and the normal 
phase when the ``heavy'' species is in excess. Although the topology of the 
phase diagram is quantitatively the same, two significant differences appear. 
The first is that, with the electron-hole bilayer, the FFLO-SF phase boundary 
on the ``heavy'' species side penetrates further into the BEC regime than in 
the ultracold atomic gas. The second difference is that, with the 
electron-hole bilayer, the FFLO region existed from the normal phase to $m=0$, 
and no phase separation between FFLO and SF was seen, except for the deep in 
the BEC regime. However in the ultracold atomic gas, phase separation of the 
SF was seen into a balanced SF and a FFLO phase. Both of these 
differences indicate that, with the electron-hole bilayer, the FFLO phase was more 
stable relative to the SF than in the ultracold atomic gas. This could be due 
to the long-range forces that act in the electron-hole bilayer whereas the 
ultracold atomic gas experiences only contact forces that would favor 
formation of tightly-bound BEC pairs.

\section{Harmonically trapped system}\label{sec:FFLOHarmonicallyTrappedSystem}

Finally, focusing on applications to ultracold atomic gases, we now address the
influence of the trap geometry on the phase behavior. Here we make use of the
local density approximation in which the chemical potential of both species,
$\mu_{\sigma}(\vec{R})=\mu_{\sigma}-V(\vec{R})$, are renormalized by the same
local trapping potential $V(\vec{R})$, the chemical potential difference, $h$,
remains fixed across the trap. Moreover, we further assume that the spatial
coordinates are rescaled to ensure a spherically symmetric trapping potential,
$V(\vec{R})=\omega R^{2}/2$. Although there is some experimental evidence
\cite{06plklh01,06pllhhs11} that the local density approximation might not be
valid \cite{06dm05,06ibld11} in some setups, we believe that its application
here will correctly address the qualitative phase structure.

\begin{figure*}
 \centerline{\resizebox{\linewidth}{!}{\includegraphics{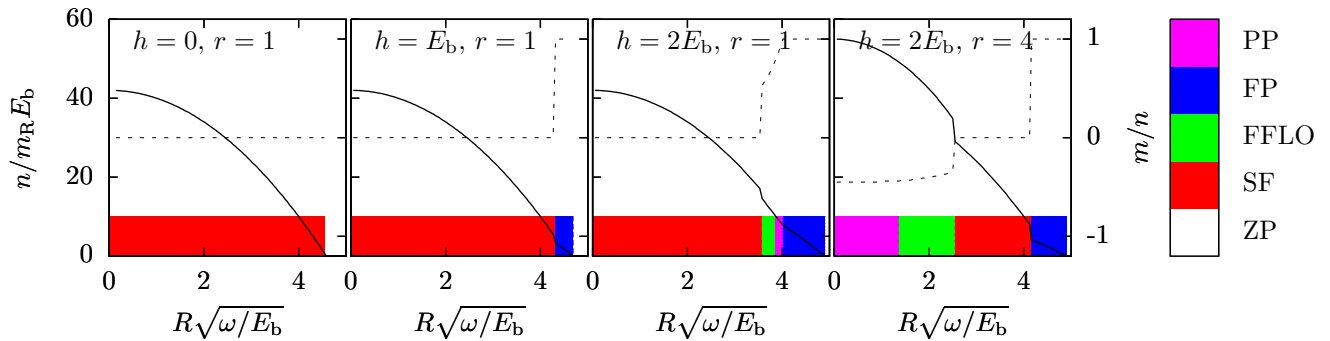}}}
 \caption{(Color online) Radial density profiles of four systems in identical harmonic traps
with different values of population imbalance, $h$, and mass ratio, $r$. The 
solid line shows the radial density based on the primary $y$-axis. The dashed 
line shows the local population imbalance based on the secondary $y$-axis. 
The band shows where different phases exist with colors labeled on the 
right-hand side of the figure. With changing radial coordinate, the separate
panels are associated with the trajectories highlighted in the 
$(h/E_{\text{b}},\mu/E_{\text{b}})$ phase diagram shown in 
\figref{fig:OmegaPhase}.}
 \label{fig:FFLOTrappedFree}
\end{figure*}

To identify the phases present, one may consider a trajectory of changing 
$\mu$ with constant $h$ and $r$ in the phase diagram of fixed chemical 
potentials. To find the total magnetization and number of particles in the 
trap, one may make use of the local relations $m=-\diffd\Omega/\diffd h$ 
(\eqnref{eqn:FFLOmagnetisation}) and $n=-\diffd\Omega/\diffd\mu$ 
(\eqnref{eqn:FFLOnumberdensity}) respectively, and then integrate over the 
trap. All trajectories will end up, at large enough radius, in the ZP regime, 
which is the edge of the particle distribution.

The profiles in four sample traps are shown in \figref{fig:FFLOTrappedFree}, 
which follow trajectories highlighted in \figref{fig:OmegaPhase}.  
The first three have species with equal masses, $r=1$. At zero population 
imbalance only the SF state is observed. With a population 
imbalance, firstly there is a central balanced SF region surrounded by a ring 
of FP majority spin particles. On increasing the population imbalance yet further, 
between the ring of FP particles and the central SF, an FFLO 
instability adjacent to a PP region is seen. The first order transition 
between the SF and FFLO region (and FP state) leads to a discontinuity in density 
and polarization. The second order transitions between FFLO, PP and FP 
states have continuous variation of density and polarizability but 
discontinuous changes in their gradients.

The final profile in \figref{fig:FFLOTrappedFree} is at an unequal mass 
ratio, $r=4$. The inclusion of mass imbalance causes the SF region in 
\figref{fig:OmegaPhase} to be biased towards the ``lighter'' species. This 
means that it is possible to have a ring of superfluidity remote from the trap 
center, or an isolated ring of FFLO instability not at the center and no 
accompanying SF region. When there are two rings of normal phase bounding the 
SF they may either both be the ``heavy'' particle normal phase if we are 
crossing the extrusion of the FFLO phase, or alternatively one might be 
``light and the other ``heavy'' if traversing right across the skewed SF 
phase. In the latter case, shown in \figref{fig:FFLOTrappedFree}, the species 
favored by the chemical potential shift dominates at the outside of the 
trap. At the center of the trap, the normal state is of the ``heavy'' species 
as superfluidity favors the ``lighter'' species.

\section{Conclusions}\label{sec:FFLOConclusions}

We have derived an analytic expression for the thermodynamic potential of a 
two-dimensional two-component atomic Fermi gas in the mean-field approximation 
with population imbalance and general mass ratio at zero temperature. A 
complementary Ginzburg-Landau analysis was used to examine the PP-FFLO transition. 
Analytical expressions for the phase boundaries separating normal and 
inhomogeneous superfluid phases have been obtained while the properties of 
the FFLO phase have been addressed numerically. Within the mean-field 
approximation, the SF phase does not sustain a population imbalance. The 
region of FFLO instability exhibits a second order phase transition from the 
PP normal phase, and first order phase transition into the balanced SF. In 
the BCS limit, a small population imbalance is sufficient to destroy pairing. 
In the BEC limit, there is phase separation between tightly-bound molecules 
and a FP normal phase. If there is a mass imbalance, the SF phase is favored 
if the majority particles are the ``lighter'' species, while the polarized 
normal state is favored if the ``heavy'' species are in excess.

A trapped geometry leads to a rich range of possible 
density profiles. If there is no mass imbalance, a SF phase is seen at the 
trap center surrounded by a PP followed by a FP normal phase of the 
majority spin species. If there is mass imbalance, then a ring of the SF and/or 
the FFLO state could be seen bordered both inside and outside by either 
species of normal phase particles.

\acknowledgements
We are grateful to Francesca Marchetti for useful discussions. The authors
acknowledge the financial support of EPSRC.

\bibliography{./references}

\end{document}